# Docking positrophilic electrons into molecular attractive potential of fluorinated methanes

Xiaoguang Ma[†,1,2] and Feng Wang[*,1]

[1]*eChemistry Laboratory, Faculty of Life and Social Sciences, Swinburne University of Technology, PO Box 218, Hawthorn, Victoria 3122, Australia*

[2]*School of Physics and Optoelectronic Engineering, Ludong University, Yantai, Shandong 264025, PR China*

The present study shows that the positrophilic electrons of a molecule dock into the positron attractive potential region in the annihilation process under the plane-wave approximation. The positron-electron annihilation processes of both polar and non-polar fluorinated methanes ($CH_{4-n}F_n$, n=0, 1, …, 4) are studied under this role. The predicted gamma-ray spectra of these fluorinated methanes agree well with the experiments. It further indicates that the positrophilic electrons of a molecule docking at the negative end of a bond dipole are independent from the molecular dipole moment in the annihilation process.

*Submitted to Journal of the Physical Society of Japan*

---

[†] *Corresponding author. E-mail: hsiaoguangma@188.com (X. G. Ma).*
[*] *Corresponding author. E-mail: fwang@swin.edu.au (F. Wang).*

# 1. Introduction

Recent research has suggested that the negative end of a polar molecule plays a primary role in positron binding and annihilation process.[1] In earlier experiments, positrons were found to annihilate predominantly with the halogen or halogen ions in polar molecules such as alkali halides.[2,3] An experimental review on the positron annihilation process has been given for both non-polar and polar molecules.[4]

It is known that the core electrons of atoms or molecules only play a minor role in the annihilation process.[5,6] However, it is still unclear why the positron annihilates the valence electrons in a molecule differently, and which valence electrons of a molecule are most likely to be annihilated by the positron. A number of available experimental measurements enable us to develop theoretical models to reveal the relations between the annihilation probability and molecular properties, such as the attractive potential, the dipole moment and the electron density distributions, etc.

Recent studies revealed that some specific valence electrons, i.e., positrophilic electrons in non-polar molecules, such as methane[7] and hexane[8], dominate the annihilation spectra. For example, the $2a_1$ electrons dominate the annihilation process of methane while the $1t_2$ electrons only play a minor role.[7]. These positrophilic electrons, i.e., the $2a_1$ electrons in methane, show that the positron annihilation probability is related to the electron density distributions and the attractive potential (negative region) of molecule. It is further found that the positrophilic electrons predominately locate on the lowest occupied valence orbital (LOVO) of the molecule as shown in the gamma-ray spectra of n-hexane as well as methane.[8] However, both methane and hexane molecules are non-polar alkanes without permanent dipole moment. Would other molecules such as polar molecules behave similarly to these alkanes, in which their gamma-ray spectra are dominated by positrophilic electrons on the LOVO orbitals?

The study of the gamma-ray spectra of positron-electron annihilation of fluorinated methanes may provide answers to the question. The series contains both non-polar, e.g., $CH_4$ and $CF_4$, and polar, such as, $CH_3F$, $CH_2F_2$, $CHF_3$ molecules. The series of fluorinated methanes have been studied experimentally with a completed set of experimental data available.[4] As a result, the fluorinated methanes are employed in the present study to further test the model of positrophilic and the LOVO electrons applied in alkane is also applicable to polar molecules. In addition, the information obtained from the fluorinated methanes can also enhance our understanding of the annihilation processes of molecules.

The theoretical expressions of the gamma-ray spectra in the positron-electron annihilation process, the local molecular attractive potential, the bond dipole moment, and the Hirshfeld charge are defined in the following section. These quantities are helpful to identify the properties which dominate the annihilation probability. Section 3 reports the theoretical results and discussions. Conclusions and summary are given in Section 4.

## 2. Theoretical methods and computations

In molecular orbital scheme, the electron or the positron wavefunction in the $i$th orbital can be expanded by Gaussian Type Functions (GTFs)

$$\psi_i(\mathbf{r}) = \sum_{klmn} C^i_{klmn} x^k y^l z^m \exp(-\alpha_n \mathbf{r}^2).$$ (1)

In this equation, $C^i_{klmn}$ are $i$th molecular orbital expansion coefficients obtained by self-consistent methods and $x^k y^l z^m \exp(-\alpha_n \mathbf{r}^2)$ are the basis functions. The particle wavefunctions are directly mapped into the momentum space[9]

$$A_{i\mathbf{k}}(\mathbf{P}) = \int \psi_i(\mathbf{r}) \varphi_{\mathbf{k}}(\mathbf{r}) e^{-i\mathbf{P}\cdot\mathbf{r}} d\mathbf{r}.$$ (2)

Where $\psi_i(\mathbf{r})$ is the wavefunction of the electron in orbital $i$ of the target in the ground state, $\varphi_{\mathbf{k}}(\mathbf{r})$ is the wavefunction of the incident positron with momentum $\mathbf{k}$, and $\mathbf{P}$ is the total momentum of the annihilation photons.

The probability distribution function of the photon momentum $\mathbf{P}$ in two-photon annihilation is then given by[6]

$$W_i(\mathbf{P}) = \pi r_0^2 c \left| A_{i\mathbf{k}}(\mathbf{P}) \right|^2$$ (3)

where $r_0$ is the classical electron radius, $c$ is the light speed. The spherically averaged $\gamma-$ray spectra for each type of electrons are then calculated by using general equations[6]

$$w_i(\varepsilon) = \frac{1}{c} \int \int_{2|\varepsilon|/c}^{\infty} W_i(\mathbf{P}) \frac{P dP d\Omega_{\mathbf{P}}}{(2\pi)^3}.$$ (4)

At large separations between the molecule and positron, it is a good approximation that the positron is described by a plane wave, and for the low positron momentum $\mathbf{k}$ in the vicinity of the target $\varphi_{\mathbf{k}} \approx e^{i\mathbf{k}\cdot\mathbf{r}} \approx 1$, which we term the low-energy plane-wave-positron approximation (LEPWP).[6] In this approximation, the influence of positron can be neglected and two quantities are usually used to reflect the molecular field effect on the positron wavefunction: the total Electrostatic Potential (ESP)[7,8] and the electric dipole moment of the molecules.[1]

The total ESP of a molecule, which represents electrostatic Coulomb interactions with a positive unit charge (the positron) in a molecule, is considered as an indicator of the positrophilic or electrophilic region of the molecule in the annihilation processes.[7,8] In the present study, it is proposed that the positron annihilating process with a molecule is described by the Local Molecular Attraction Potential (LMAP),

$$U(\mathbf{r}) = -\rho_{\mathrm{p}}(\mathbf{r}) \cdot V_{\mathrm{mol}}(\mathbf{r}). \tag{5}$$

Where the $\rho_{\mathrm{p}}(\mathbf{r})$ is the positron density and $V_{\mathrm{mol}}(\mathbf{r})$ is the total ESP of the molecule, which consists of the nuclear $V_{\mathrm{nuc}}(\mathbf{r})$ and electrons $V_{\mathrm{ele}}(\mathbf{r})$ in a molecule

$$V_{\mathrm{mol}}(\mathbf{r}) = V_{\mathrm{nuc}}(\mathbf{r}) + V_{\mathrm{ele}}(\mathbf{r}) = \sum_A \frac{Z_A}{|\mathbf{r} - \mathbf{R_A}|} - \int \frac{\rho_{\mathrm{e}}(\mathbf{r}')}{|\mathbf{r} - \mathbf{r}'|} d\mathbf{r}'. \tag{6}$$

Under the plane-wave approximation, the positron density has little effect on the LMAP due to $\rho_{\mathrm{p}}(\mathbf{r}) = |\varphi_{\mathbf{k}}| \approx 1$, As a result, the annihilating process of a positron is totally dependent on the $V_{\mathrm{mol}}(\mathbf{r})$ under this approximation.

Furthermore, the Hirshfeld charge scheme is also employed to understand the site selectivity in the present study. The Hirshfeld charges provide information of the electronegativity of a molecule, which can help to understand the likely positron attraction sites in a molecule.[10,11] The Hirshfeld charge is defined by[12]

$$Q_{\mathrm{A}}^{\mathrm{H}} = Z_A - \int \frac{\rho_{\mathrm{A}}(\mathbf{r})}{\rho_{\mathrm{pro}}(\mathbf{r})} \rho_{\mathrm{mol}}(\mathbf{r}) d^3\mathbf{r} \tag{7}$$

Where $Z_{\mathrm{A}}$ is the nuclear charge, and $\rho_{\mathrm{A}}(\mathbf{r})$ is the spherically-averaged atomic electron density centered on nucleus A. The $\rho_{\mathrm{pro}}(\mathbf{r})$ and $\rho_{\mathrm{mol}}(\mathbf{r})$ are the sums of electron density over the atoms belonging to the promolecule and the molecule, respectively.

The theoretical calculations are mainly implemented by using the Gaussian09 computational chemistry package. In the *ab initio* Hartree-Fock calculations, the TZVP basis set is used, i.e., the calculation model is the HF/TZVP model. In the TZVP basis set, atomic carbon and fluorine orbitals are constructed by the *5s9p6d* scheme of Gaussian type functions (GTFs), while atomic hydrogen orbitals are constructed by the *3s3p* scheme of GTFs. The molecular wavefunctions are then obtained using the HF/TZVP model and based on the optimised structural parameters using the same model. This HF/TZVP model is used for the gamma-ray spectra studies of methane[7] and hexane[8] molecules for consistency and comparison purposes.

## 3. Results and discussions

The structures of the fluorinated methane molecules together with the calculated total electron density mapped on the ESPs and Hirshfeld charge using the HF/TZVP model are shown in Fig.1. Under the plane-wave approximation, the LMAP in Eq.(5) becomes the effective ESP in an molecule for a positron. The negative region(s) of the ESP(coloured in red in Fig.1) represents the positive LMAP, $U(r)$, in Eq.(5). It is this region of the LMAP which attracts the positron and serves as an indicator of the annihilation probability. For instance, in the methyl fluoride $CH_3F$ molecule, the negative potential (red) concentrates on the fluoride atom F which formes a hemisphere as shown in Fig.1(a). Hence the region in Fig.1(a) colored in red is the effective ESP, i.e., the LMAP is the strongest region among all other fluorinated methane molecules in this figure. This indicates that there is a strong attractive force for the coming positron to the fluoride atom of $CH_3F$.

The Hirshfeld charge calculated using Eq.(7) also shows a similar variation with the above ESP as shown in Fig.1. According to the Hirshfeld charge distribution calculated using Eq.(7), the net charge around hydrogen atom is almost the same and decreases very lightly from $CH_4$ which can be neglected at all. Except that, the Hirshfeld charges on F and C show the increase from $CH_4$ or $CH_3F$ to $CF_4$. The fluoride atoms are negatively charged, but the hydrogen and carbon atoms always exhibit positive charges in the fluorinated methanes. The value of $Q^H$ of the fluorine atoms increases from -0.310 a.u. to -0.208 a.u. from $CH_3F$ to $CF_4$ as more F atoms bond to the central carbon atom. Apparently, the fluorine atoms play a role as positron acceptors. This is in agreement with the findings that the positrons annihilate predominantly with the negative atoms in molecules[2].

Under the thermodynamic equilibrium state (or with a low energy), a positron approaches the target electrons towards the negative potential region in the direction that it feels a maximum local attractive force of a molecule.[7,8] The electrons trapped into this region are more likely to annihilate with the positron. Furthermore, the extra electrons (negative Hirshfeld charge) located around F atom are sufficient for the annihiliation. For these four fluorinated methanes as shown in Fig.1, the electrons of the specific orbitals distributed most of the electrons around fluorine atoms dominate the annihilation process. These electrons are the phositrophilic electrons of the molecules[7,8].

Fig.2 shows the positrophilic electrons of $CH_3F$ molecule in positron-electron annihilation process. The region in Fig.2(a) coloured red which locates around the fluoride atom of $CH_3F$ represents the positrophilic region (negative end) of the molecule. Fig.2 (b) shows the contour color-filled map of electron density distribution of the $3a'$ electrons in the F-C-F plane in the ground electronic states of $CH_3F$. The $3a'$ electrons are on the lowest occupied valence orbitals (LOVO)[8] of $CH_3F$. As seen in Fig2(b), the $3a'$ electrons dock into the LMAP region as indicated in Fig.2(a). As a result, the $3a'$ electrons likely dominate the annihilation process of the $CH_3F$ molecule.

Apart from the LOVO electrons, other valence electrons can also contribute to positrophilic electrons of a molecule.[8] Fig.2(c) shows the contour color-filled map of the electron density distribution of the $4a'$ electrons of $CH_3F$ The superposition of the electrons in the $3a'$ and $4a'$ orbitals is given in Fig.2(d), which exhibits almost the same distribution of the negative ESP (the positive LMAP). That is, the shape of the superpositioned positrophilic electrons fits well (docks) into the shape of the negative ESP, which contains the most probable electrons to annihilate with positron. As a result, the $3a'$ and $4a'$ electrons of $CH_3F$ are the positrophilic electrons, which are determined by the docking between Fig.2(a) and Fig.2(d). In other words, the distribution of the negative potential is dominated by the distribution of the $3a'$ and $4a'$ electron densities and therefore, these positrophilic electrons dominate the annihilation of $CH_3F$.

Fig.3 shows the results of the simulated gamma-ray spectra of methyl fluoride $CH_3F$ molecule in positron-electron annihilation process, in comparison with two-Gaussian fitted experimental results in gas phase.[4] As only the valence electron contributions are important in the annihilation process[1], only the gamma-ray spectra of the dominant valence electrons and the total valence electrons are given. As shown in Fig.3, the contributions to the profiles of $\gamma-$ray spectra are orbital dependent, as orbitals contain the information of the distribution of the electron densities. The $3a'$ electrons, i.e., the LOVO electrons result in 2.91 KeV of full width at half maximum (FWHM) which is slightly higher than measurement of 2.68 KeV. While the $4a'$ electrons give a 2.54 KeV of FWHM which is less than the same experiment. Superposition of the electron densities in the orbitals reproduces the negative potential ESP, as the positrophilic electrons $(3a'+4a')$ of $CH_3F$ gives 2.76 KeV of FWHM, which agrees very well with the measured one (i.e., 2.68 KeV).[4] However, superposition of all valence electrons of $CH_3F$ gives 4.32 KeV of FWHM, which is significantly different from experimental result of 2.68 KeV for the same molecule. This indicates that not all the valence electrons make the same contributions to this polar molecule, in agreement to the previous findings in alkanes.[7,8] The positrophilic electrons of $3a'+4a'$ dominate the annihilation process in $CH_3F$ molecule and the annihilation process takes place at the electrons in the negative ends of molecule.

The docking correlation between the positrophilic electrons and the ESPs of the $CH_2F_2$, $CHF_3$ and $CF_4$ molecules is presented Fig 4. Panel (a) of each molelcule gives the LOVO electrons which dominate the positrophilic electrons of the moelcules. Panel (b) represents minor contributions to the positrophilic electrons of the same molecules. The total positrophilic electrons distributions of the molecules are shown in Panel (c). As one can see that the densities of the positrophilic electrons (c) dock into the corresponding negative regions of the ESP (i.e., the positive LMAP) in Panel (d) for each molecule. For example, the positrophilic $4a'$ and $6a'$ electrons of $CH_2F_2$ have the dominant probability to annihiliate with the incoming positron. The positrophilic electrons, $3a_1+2e$ of $CHF_3$ and $3a_1+2t_2$ of $CF_4$, dock into the negative potential, respectively. When the shape of the negative potential of a molecule fits the shape of the electron densities of a combination of certain electrons, it is called the electrons dock into the potential well. As a result, the positron annihilates these positrophilic electrons of the molecule spontaneously.

The gamma-ray spectra for the remaining fluorinated methanes, that is, $CH_2F_2$, $CHF_3$ and $CF_4$ are shown in Fig.5. The width of the experimental $\gamma-ray$ spectra of $CH_2F_2$ is compared with simulated widths under individual contributions. For example, the experimental FWHM of $CH_2F_2$ is given by 2.76 KeV, which is the same as the FWHM of $CHF_3$ but lager than the FWHM of $CH_3F$ (2.68 KeV) . The spectra of the positrophilic electrons are also wider than the corresponding one of the $CH_3F$ molecule. For the LOVO electrons, the width is the lowest among all the spectra of $CH_2F_2$. While in the $CH_3F$ molecule, the lowest one is the $4a'$ electron which is not the LOVO electron. The width of LOVO is 2.39 KeV, and $6a'$ is 3.28 KeV for $CH_2F_2$. The positrophilic electron gives out 2.83 KeV which agrees well with the experiments. Furthermore, by comparison of these two molecules, the total valence spectra disagree largely with the experiments respectively.

In comparison, The $CH_2F_2$ and the $CHF_3$ have the same experimental width[4] of the spectra. The width of the positrophilic electrons is 2.98 KeV for $CHF_3$, bigger than the corresponding one of $CH_2F_2$. The positrophilic electrons are the $3a_1$ and $2e$ electrons. The $3a_1$ is the LOVO and also the lowest width among all the spectra of $\gamma-ray$ with more separation from the experiments. The $2e$ electrons have wider spectra which make the positrophilic electron spectra not agreeable very well with the experiments. In particular notably, the positrophilic electrons give out the spectra width of 2.98 KeV for $CHF_3$ while the width of the positrophilic electrons of $CH_2F_2$ is only 2.83 KeV. The difference is about 0.15 KeV. However, the experimental width almost the same, 2.76 KeV. This can not be explained by the positrophilic priciple in the present study and might be beyond the plane-wave approximation.

The gamma-ray spectra of carbon tetrafluoride molecule $CF_4$ in positron-electron annihilation process are compared with two-Gaussian fitted experiments.[4] For each individual orbital, the $\gamma-ray$ spectra of $3a_1$ and $2t_2$ electrons agree with experiments more than those electrons occupying in the other orbitals. However, $\gamma-ray$ spectrum of each orbital does not agree very well with the experiments. The $3a_1$ electron presents 2.01 KeV which is less than experiment 2.96 KeV. While the $2t_2$ electron gives a 3.80 keV of FWHM which is bigger than the same experiment. According to positrophilic principle in annihilating process as shown in Fig.4, the most dominant contributions to the annihiliating process are from both the $3a_1$ and the $2t_2$ electrons. These $3a_1+2t_2$ electrons agree very well with experiment about the $\gamma-ray$ spectra.

The other orbitals or the other summation schemes will not give the same symmetry and distribution of the negative potential, so these electrons will contribute less to the annihilating processes.

## 4. Conclusions

In summary, the present study demonstrates to determine the positrophilic electrons of a molecule from docking into the positron attractive potential region in the annihilation process. A concept of local molecular attractive potential is presented to accommodate the docking in the positron-electron annihilation process. The positrophilic electrons of a molecule determined from the docking method are dominated by the LOVO electrons of the fluorinated methane molecules, regardless if it is a polar molecule in agreement with our previous findings of alkanes.[7,8] The present study further indicates that the negative end of the molecular permanent or induced dipole moment is an important property in the annihilation process of a molecule. The predicted gamma-ray spectra of these fluorinated methanes agree reasonably with the experiments. More evidences on positrophilic electrons docking in other molecules, such as alkanes ($C_nH_{2n+2}$) will be published elsewhere.

## 5. Acknowledgements

This project is supported by the Australia Research Council (ARC) under the discovery project (DP) scheme. National Computational Infrastructure (NCI) at the Australia National University (ANU) under the Merit Allocation Scheme (MAS) is acknowledged. Swinburne University's GPU supercomputing facilities are also acknowledged. Professor Ma acknowledges the support of the Natural Science Foundation Project of Shandong Province (No.ZR2011AM010).

# Figure captions

Figure 1: The total electron density distributions mapped on the total electrostatic potentials (ESP) of the methyl fluoride $CH_3F$, difluoromethane $CH_2F_2$, trifluoromethane $CHF_3$, and carbon tetrafluoride $CF_4$ molecules, respectively.

Figure 2: The totla electrostatic potential is shown in (a) red represents the negative value while blue represents the positron one. The contour color-filled map of electron density distribution of $CH_3F$ in arbitary F-C-H plane shown in (b): $3a'$; (c): $4a'$ and (d): $3a' + 4a'$.

Figure 3: Gamma-ray spectra of $CH_3F$ in positron-electron annihilation process.

Figure 4: The contour color-filled map of electron density distribution for $CH_2F_2$ ( (a): $4a'$, (b): $6a'$ and (c): $4a' + 6a'$), $CHF_3$ ((a): $3a_1$, (b): $2e$ and (c): $3a_1 + 2e$), and $CF_4$ ((a): $3a_1$, (b): $2t_2$, and (c): $3a_1 + 2t_2$) molecules.

Figure 5: Gamma-ray spectra of $CH_2F_2$, $CHF_3$, and $CF_4$ molecules.

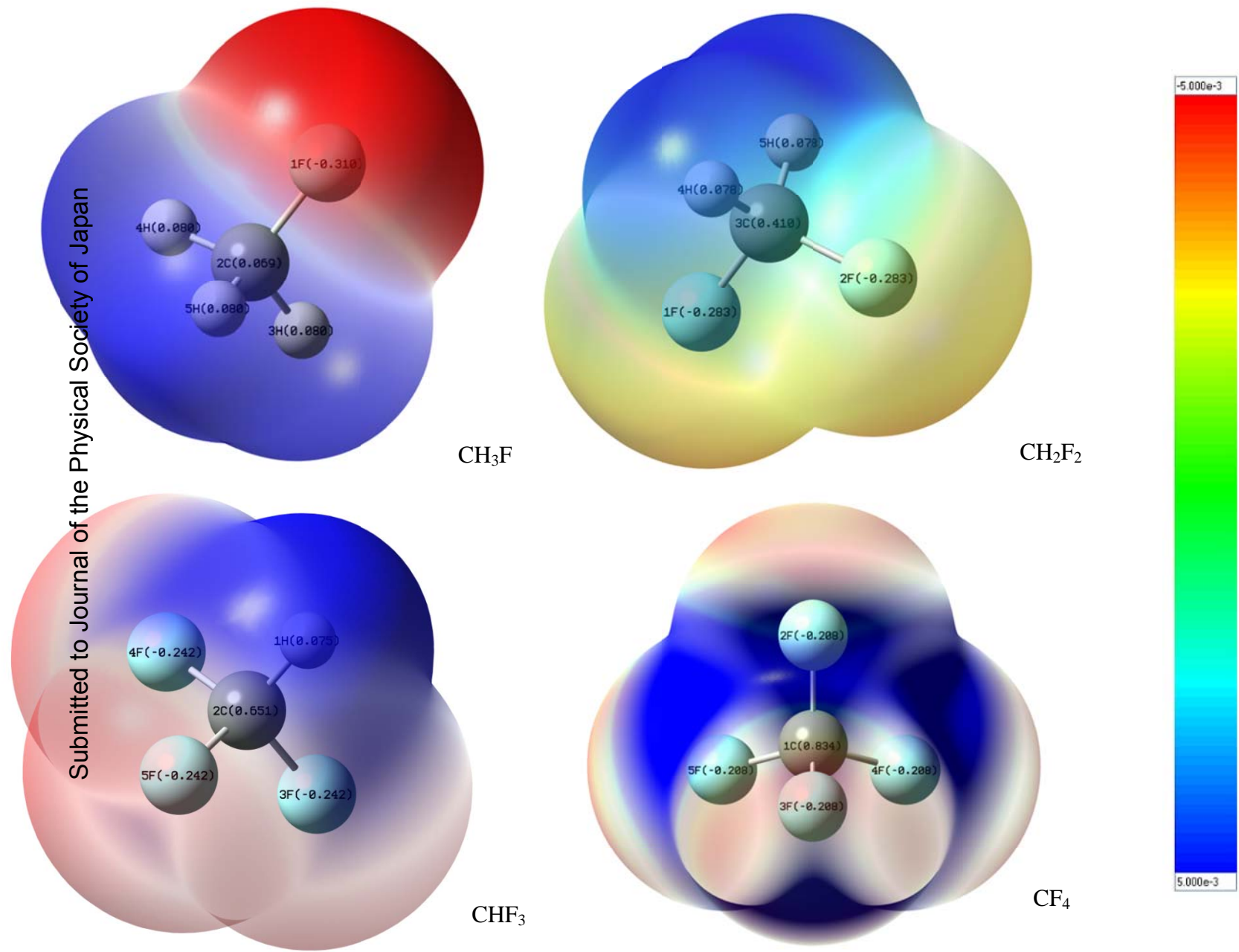

CH₃F

CH₂F₂

CHF₃

CF₄

Fig.1

CH₃F

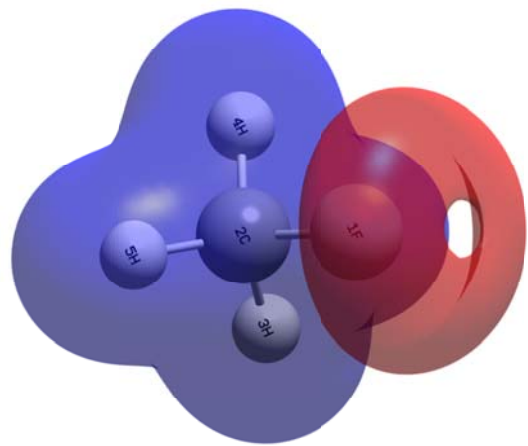

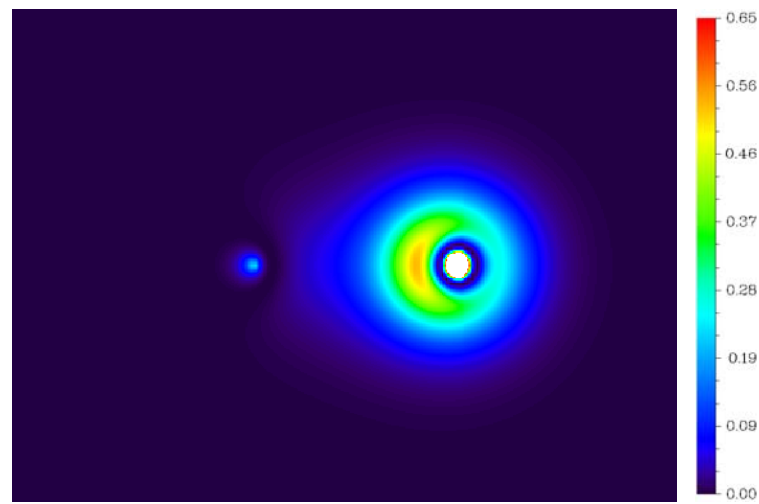

(a)

(b)

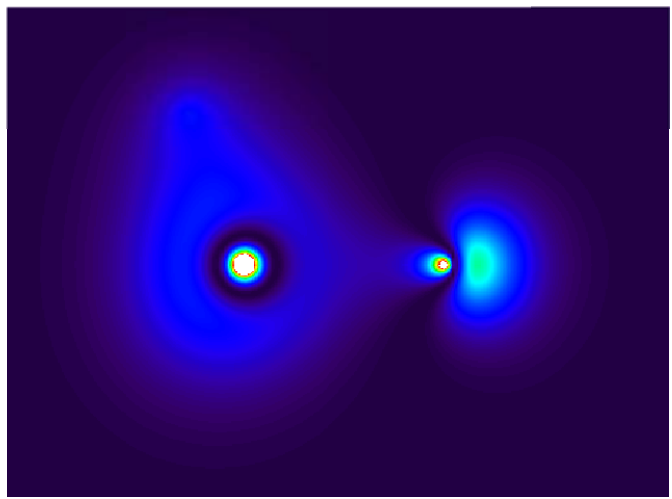

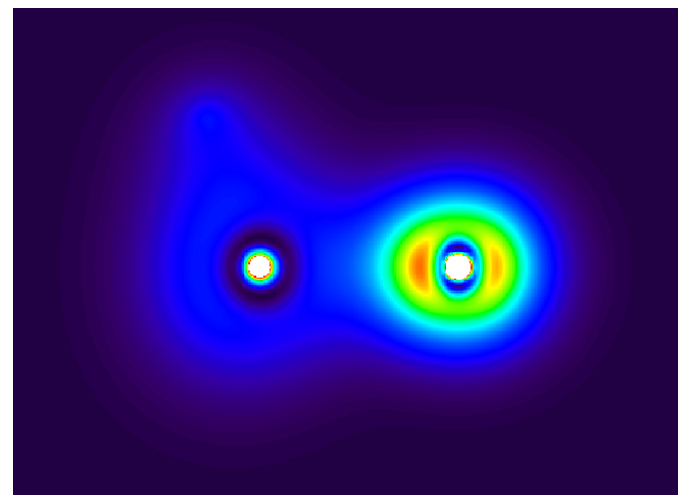

(c)

(d)

Fig.2

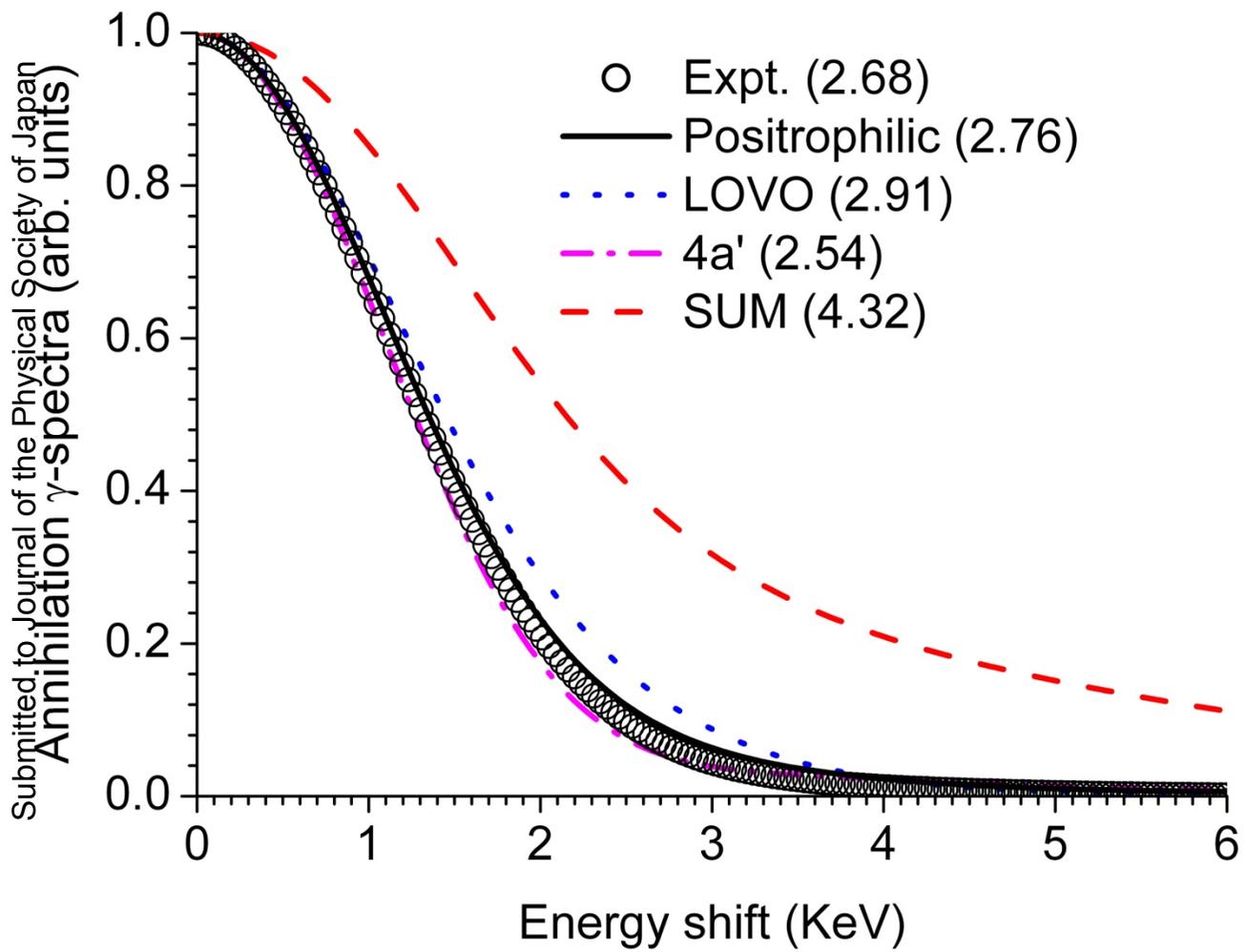

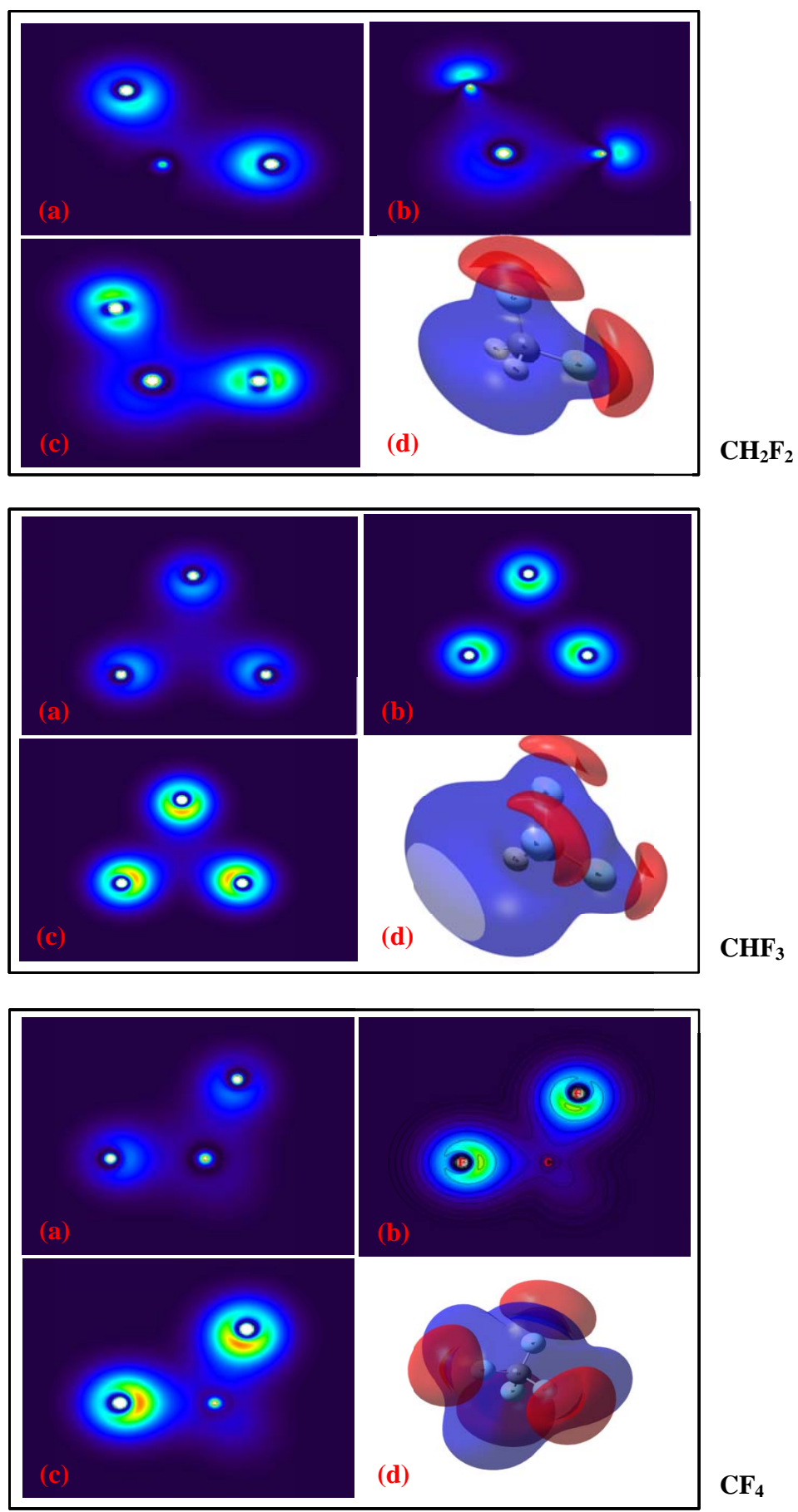

(a)
(b)
(c)
(d) $CH_2F_2$

(a)
(b)
(c)
(d) $CHF_3$

(a)
(b)
(c)
(d) $CF_4$

Fig.4

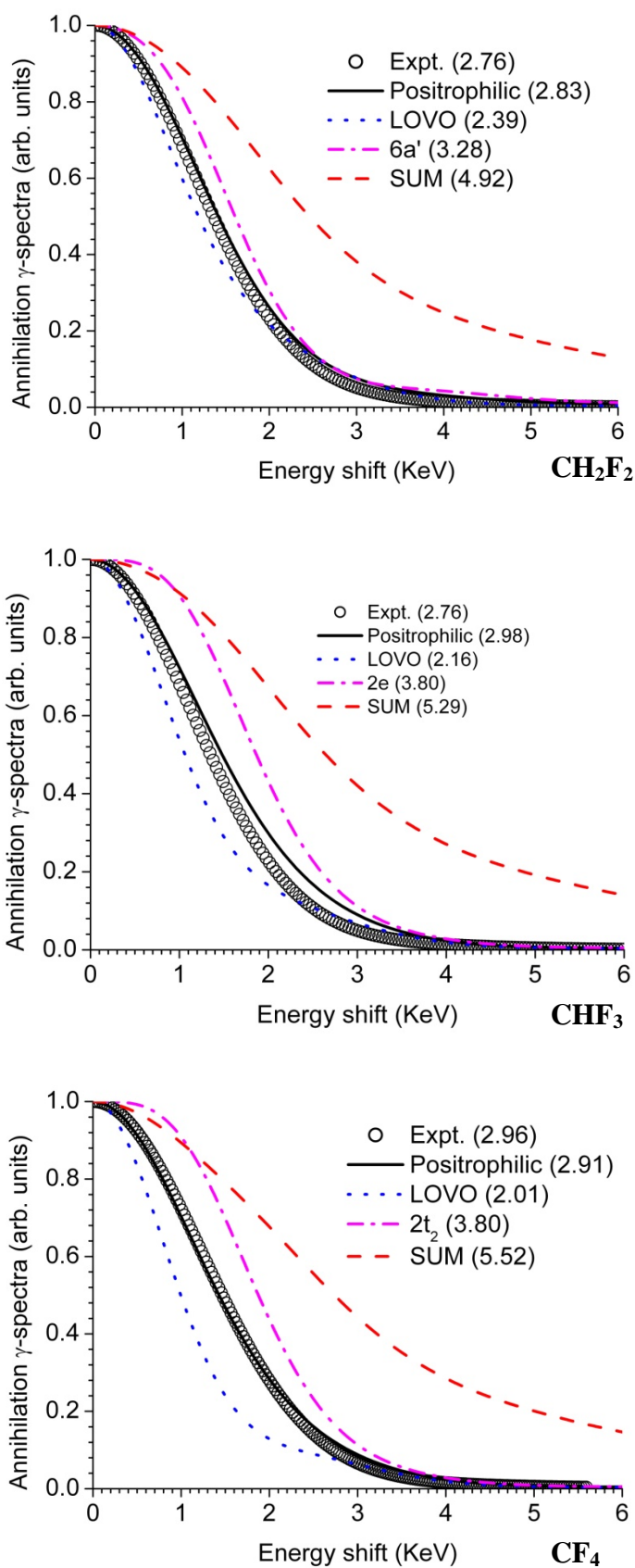